\documentclass{iau} 
\usepackage{graphicx, natbib}

\title[Probing NS interiors with pulsar glitches] %% give here short title %%
{Probing neutron star interiors with \\ pulsar glitches}

\author[Brynmor Haskell]   %% give here short author list %%
{Brynmor Haskell}

\affiliation{Nicolaus Copernicus Astronomical Center of the Polish Academy of Sciences\\ Ulica Bartycka 18, 00-716 Warszawa, Poland \\email: {\tt bhaskell@camk.edu.pl}}

\pubyear{2017}
\volume{337}  %% insert here IAU Symposium No.
\jname{Pulsar Astrophysics -­ The Next 50 Years}
\editors{P. Weltevrede, B.B.P. Perera, L. Levin Preston \& S. Sanidas, eds.}

\begin{document}

\maketitle

\begin{abstract}

Pulsar glitches are thought to be probes of the superfluid interior of neutron stars. These sudden jumps in frequency observed in many pulsars are generally assumed to be the macroscopic manifestation of superfluid vortex motion on a microscopic scale. Resolving and modelling such phenomena on the scale of a neutron star is, however, a challenging problem which still remains open, fifty years after the discovery of pulsars. In this article I will review recent theoretical progress, both on the microscopic level and on the macroscopic level, and discuss which constraints on the models can be provided by observations. 

\keywords{stars: neutron, pulsars: general, hydrodynamics, equation of state}
%% add here a maximum of 10 keywords, to be taken form the file <Keywords.tex>
\end{abstract}

\firstsection % if your document starts with a section,
\section{Introduction}

Pulsar glitches are sudden jumps in frequency first observed in radio pulsars soon after their discovery \citep{crab1, crab2, vela1, vela2}. Following these events an increase in the spin-down rate is often observed, and in some cases we see also an exponential relaxation towards the previous spin-down rate, over long timescales of days to months. These long timescales were, in fact, interpreted, soon after the first observations, as the signature of a large scale superfluid component in the interior of the neutron star, only loosely coupled to the `normal' component that is tracked by the electromagnetic emission \citep{Baym69}. A more quantitative explanation of the recovery was given by the superfluid vortex `creep' theory of Alpar and collaborators \citep{Alpar84a}.

Superfluids are expected in neutron star interiors, as at such high densities neutrons will pair and become superfluid soon after the birth of the star (see \citet{HasSed2017} for a recent review), and this is at the heart of most models of pulsar glitches. A superfluid rotates by forming an array of quantised vortices that carry the circulation. In the crust of the star, however, vortices can be strongly attracted (`pinned') to nuclear clusters. If this is the case the neutron superfluid cannot expel vorticity and cannot spin-down, thus lagging behind the `normal' component of the star that is being spun-down by electromagnetic emission. As the lag increases so do classical hydrodynamical lift forces (Magnus forces), which eventually become strong enough to overcome the pinning. Vortices are then free to move out, mediating a rapid exchange of angular momentum between the components and giving rise to a glitch \citep{AndItoh75}.

Although this general picture is fairly well established, 50 years after the discovery of pulsars, a full quantitative model for pulsar glitches is still elusive. In particular several possibilities exist for the trigger mechanism, with the main theories revolving around crust quakes \citep{Rud69}, hydrodynamical instabilities \citep{Kostas09} and vortex avalanches \citep{Cheng88, Lila13}. In fact, recent work on the small scale motion of superfluid vortices in a pinning potential has been fairly successful in reproducing features of the observed glitch size and waiting time distributions \citep{Lila11, Lila13} (see \citet{HasRev} for a recent review). Nevertheless scaling up these results, obtained for hundreds or thousands of vortices in a small volume, to the scale of a 10 km neutron star with  $\approx 10^{17}$ vortices is clearly a challenge. In fact, recent attempts to construct a mean field model for vortex motion on large scale have shown that the hydrodynamical response of the star can have a large impact on the observed glitch properties, which will deviate from the microphysical predictions \citep{Haskell16}.

There are thus two sides to the problem: the first regards the issue of vortex pinning, and how much angular momentum can be stored in the superfluid and released in a glitch. In the following I will discuss recent theoretical results (section \ref{pin1}) and show how observations have been used to constrain the physics of neutron star crusts and estimate masses of glitching pulsars (section \ref{masse}).
The second regards the issue of the trigger mechanism. Here I will focus on vortex avalanches (section \ref{avalanche}) and discuss attempts to include their effect in large scale hydrodynamical simulations of glitches (section \ref{large}).

\section{Vortex pinning}
\label{pin1}
 
One of the main ingredients for pulsar glitch models is the strength of the pinning interaction, which determines how much angular momentum can be exchanged between the superfluid and the normal component. Several estimates of the interaction between a vortex and a nucleus exist (e.g. \citet{AlpPin, AndPin, EB88, Seveso16, Wlaz16}). Realistic results have been recently obtained by \citet{Wlaz16}, who find a repulsive interaction between the nucleus and a vortex line. \citet{Seveso16}, on the other hand, have calculated the force per unit length acting on a vortex interacting with many pinning sites, for realistic values of the tension. These forces can be used to calculate the amount of angular momentum that can be stored in the superfluid before the Magnus force unpins vortices \citep{Pierre11, Seveso12}. In general the results indicate that a sufficient reservoir of angular momentum can be accumulated to explain the size of giant glitches, which have jumps in frequency $\Delta\nu\approx 10^{-4}$ Hz \citep{Pierre11, Has12}. The time it would take a spinning down neutron star to accumulate the maximum lag is well matched to the waiting time between giant glitches in the Vela pulsar (approximately 1000 days), and explains why pulsars that spin-down faster glitch more often (e.g. PSR J0537-6910 spins down approximately 10 times faster than the Vela pulsar, and glitches approximately every 100 days - see \citet{Has12} for a detailed discussion of several other pulsars). There is, in fact, mounting evidence that the distribution of glitch sizes is bimodal \citep{300, Fuentes}, and the mechanisms described above (the so-called `snowplow' model of \citet{Pierre11}) would provide a natural explanation for the large glitch population. We will discuss in section (\ref{avalanche}) how vortex knock-on may explain the general population of glitching pulsar and the prevalence of smaller glitches in many systems, with size distributions that are consistent with power-laws \citep{Melatos08}.

\subsection{Estimating pulsar masses}
\label{masse}

In the previous section we have simply discussed the largest glitches observed in a pulsar. It is, however, also highly instructive to consider the overall fraction of the spindown that has been reversed by glitches over the time a system was observed. This is usually described in terms of the activity $\mathcal{A}$ defined observationally as:
\begin{equation}
\mathcal{A}=\frac{1}{\tau_o}\sum_i\frac{\Delta\nu_i}{\nu_i}
\end{equation}
where $\tau_o$ is the timespan over which the system has been observed, and the index $i$ labels the various glitches. This quantity can be used to constrain the angular momentum reservoir, as one has \citep{Crust12, Chamel13}:
\begin{equation}
\frac{I_s}{I_n}\approx-\frac{\nu}{\dot{\nu}}\mathcal{A}\frac{m_n^*}{m_n}
\end{equation}
where $I_s$ is the moment of inertia of the superfluid that gives rise to the glitch, and $I_n$ is the moment of inertia of the normal component coupled during the glitch itself. The effective mass $m_n^*$ of neutrons in the crust (with bare mass $m_n$) also enters this estimate, and is very important, as calculations by \citet{entrain} find large values for the effective mass. This essentially means that the superfluid neutrons are less mobile and the amount of angular momentum available for a glitch decreases. In general to explain the activity of the Vela pulsar one needs $I_s/I_n\approx 4-6\%$, which is more than what is available in the crust unless the equation of state is very soft and the mass of the star very small (below $1 M_\odot$). Recent calculations by \citet{Watanabe17}, however, find that including pairing in the calculation of effective masses in the crust strongly reduces the values, and \citet{Newton15} have shown that mutual friction in the core also decouples part of the star during a glitch, reducing the tension and allowing for constraints on the equation of state and nuclear physics parameters, such as the slope of the symmetry energy $L$.

The analysis can be taken further, and fits to the activity can be used to `measure' the mass of a neutron star, given an equation of state. This was first done by \citet{Ho15}, assuming only S-wave pairing for the superfluid and considering the thermal evolution of the star, and by \citet{MassPierre}, who used realistic pinning profiles from \citet{Seveso16} and considered both S-wave and P-wave pairing. In particular \citet{MassPierre} also obtain a robust upper limit on the mass by fitting for the largest observed glitch in a pulsar, and constrain the mass for several equations of state. The distribution of masses they obtain is roughly in agreement with the distribution of observed masses, although none of these are measured in glitching pulsars. Future independent mass measurements for a glitching pulsar would thus be very valuable and help test and constrain the models.

\section{Triggers: vortex avalanches}
\label{avalanche}

Up to now we have discussed static models for the angular momentum reservoir available for glitches, and the `snowplow' model for giant glitches, which assumes that a critical lag is built up and released quasi-periodically. This is, indeed, what happens for the Vela pulsar, and for PSR J0537-6910, in which a correlation is also observed between the size of a glitch and the waiting time to the next \citep{Middle06, Ant17}, which is consistent with the `snowplow' model. Nevertheless these two pulsars are the only two that exhibit this kind of quasi periodic behaviour, while in all other pulsars glitch size distributions are consistent with powerlaws, and waiting times with exponentials \citep{Melatos08}. 

Such distributions are the hallmark of a self organised critical process, where stresses built up by slow external drivers (in this case the external spindown) are released by faster local interactions between neighbouring elements. In particular \citet{Lila13} have shown that vortex knock-on could lead to this situation, and vortex avalanches could thus be a trigger for glitches. Quantum mechanical simulations of pinned vortices in a spinning down container, in fact, naturally lead to power law distributions for sizes and exponentials for waiting times \citep{Lila11}. The numerical limitations of these tools, however, only allow for simulations of a limited number of vortices, relatively close together. In a realistic neutron star there are of the order of $10^{17}$ vortices, separated by up to $10^{10}$ pinning sites, which make scaling up the results a challenge. To do this it is necessary to move up in scales, and rather than resolve individual vortices on the scale of their coherence length (a few fm), treat them as massless objects subject to the pinning force, the Magnus force and dissipation. In this case one can simulate the system on the inter-vortex scale of a few millimeters,  and recent results have shown that if the system self-organises and is close enough to the unpinning threshold (to within a few percent) a vortex will still manage to skip over many pinning sites and knock on a neighbour  \citep{Haskellhop}. 

\section{Large scale models of glitches}
\label{large}

In the previous section we have seen that small scale vortex motion and interactions are an important element of the physics of glitches. However, microscopic simulations cannot be scaled up to the whole star. This poses a problem, as large scale hydrodynamical models of superfluid neutron stars generally need to average over scales larger than the inter-vortex separation to define an average rotation rate for the superfluid, thus losing information on vortex-vortex interactions. A mean field approach is thus necessary to encode these sub-grid effects in larger scale models (see e.g. \citet{Mongiovi, Fulgenzi}). 

\begin{figure}[b]
% \vspace*{-2.0 cm}
\centerline{
 \includegraphics[width=2.6in]{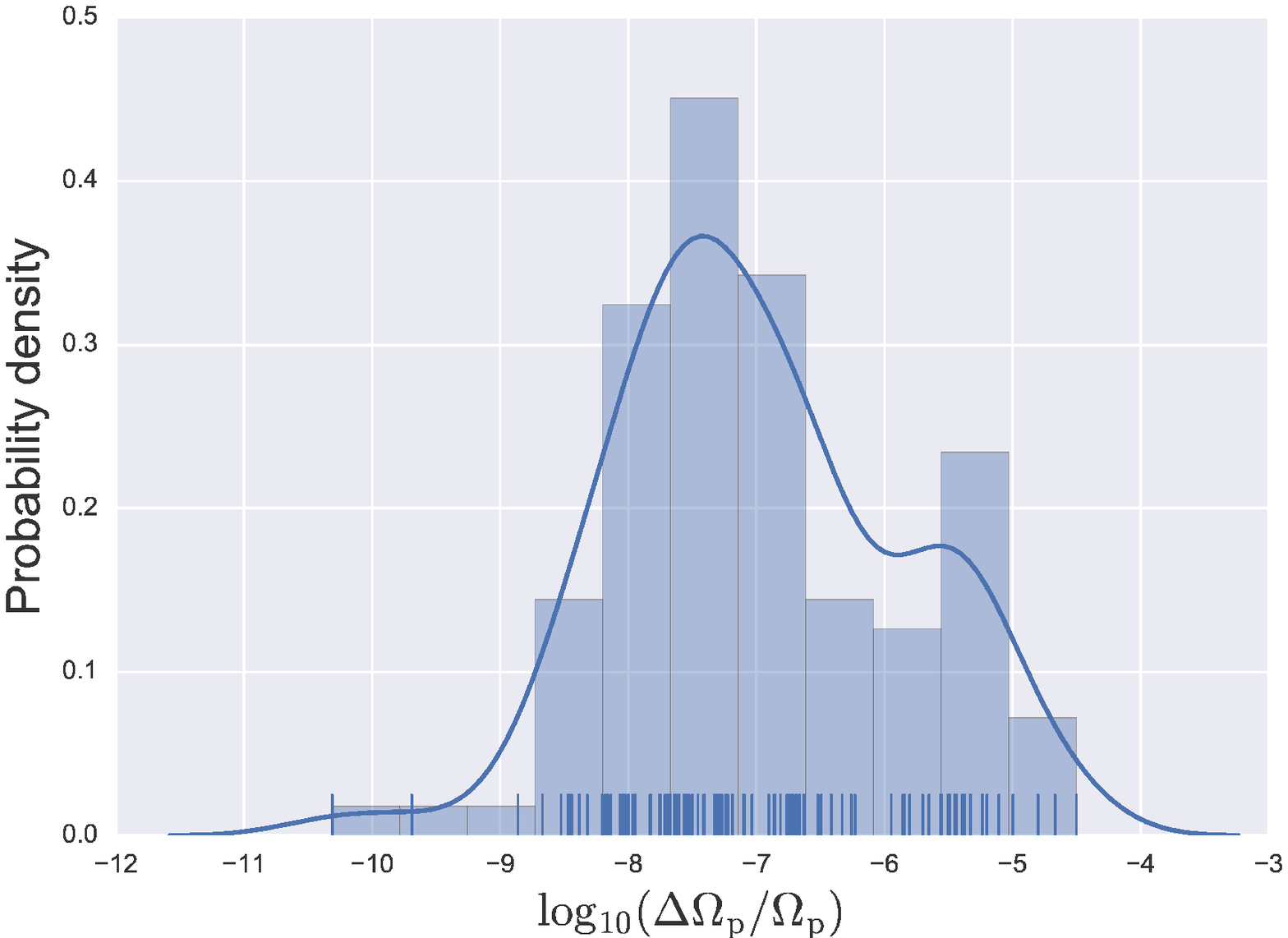} \includegraphics[width=2.6in]{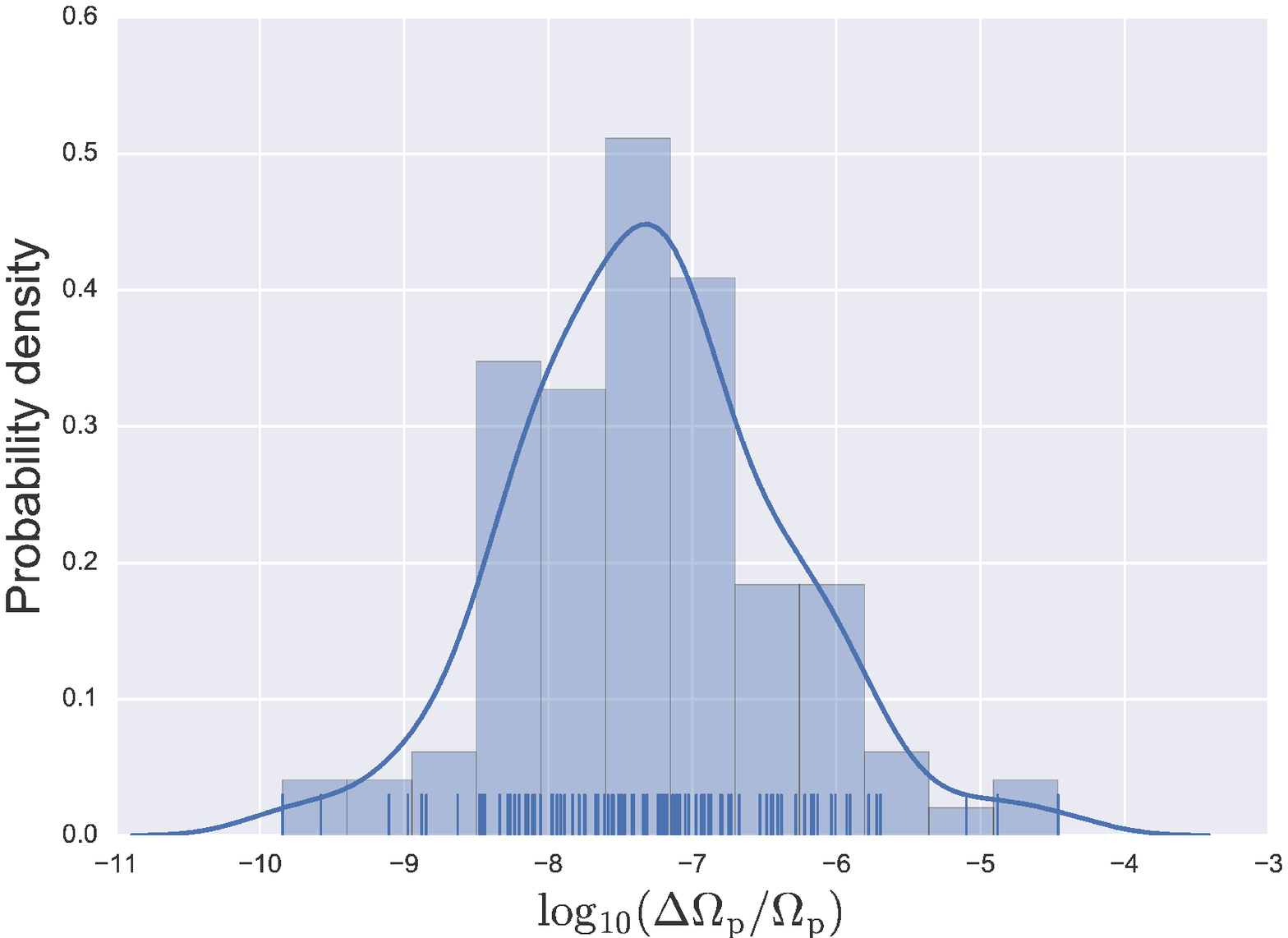} }
% \vspace*{-1.0 cm}
 \caption{Examples of probability distributions for the size $\Delta\Omega/\Omega$ of a glitch, for microscopic power-law distributions for $\gamma$, with index $n=-1.05$ on the left and $n=-1.5$ on the right, as detailed in \citet{Haskell16}. Both distributions exhibit a cutoff at small sizes, due to the fact that smaller events are also slower, and tend to appear as gradual changes in the spin-down rate, rather than abrupt spin-up events, i.e. appear more like timing noise than glitches. }
   \label{fig1}
\end{figure}

One approach is to define an additional parameter, the fraction of free vortices $\gamma=n_f/n_v$, where $n_f$ is the surface density of free vortices, and $n_v$ the total surface density of vortices. 
The simplified two-fluid evolution for the frequency of the superfluid neutrons $\Omega_n$ and normal component $\Omega_c$, neglecting entrainment (i.e. with $m^*_n=m_n$) and pinning forces for simplicity, can be written as:
\begin{eqnarray}
\dot{\Omega}_n &=&\gamma\kappa n_v \mathcal{B} (\Omega_c-\Omega_n)\label{eom1}\\
\dot{\Omega}_c &=&- \frac{\rho_n}{\rho_c}\gamma\kappa n_v \mathcal{B} (\Omega_c-\Omega_n)-T_{sd} \label{eom2}
\end{eqnarray}
where $\rho_n$ and $\rho_c$ are the superfluid neutron and normal fluid densities, $T_{sd}$ is the external spin-down torque, $\kappa$ the quantum of circulation and $\mathcal{B}$ the mutual friction parameter that encodes the strength of the superfluid drag.
Assuming that $\gamma$ is a random variable drawn from micro-physically motivated distributions we have shown that the coupling of vortices to the fluid has a significant impact and the macroscopic distributions for glitches are not necessarily power-laws, but show low-amplitude cutoffs \citep{Haskell16}. This can be seen in the example in Figure \ref{fig1} and is in agreement with the cutoff for small sizes that is observed in the Crab pulsar \citep{EspMin}. Furthermore different values of $\gamma$ strongly influence the relaxation after the glitch, with large values typically associated with exponential relaxations, and smaller values with simple steps in frequency and/or frequency derivative \citep{HasAnt}. Current work is focussed on testing prescriptions for the evolution of $\gamma$ in order to model vortex avalanches in a hydrodynamical framework (Khomenko \& Haskell, in preparation).

\section{Conclusions}

Fifty years after the discovery of pulsars, a quantitative model for pulsar glitches is still elusive. This is mainly due to the huge range in scales that is needed to model the phenomenon in a neutron star. Models need to range from the Fermi scale of the vortex-cluster interaction responsible for pinning, to the millimetre scale of vortex-vortex interactions that could trigger avalanches and glitches, and up to the kilometre scale of the star's observable response.

In this article I have briefly reviewed recent progress in our understanding of the pinning forces, and shown how observations of giant glitches, and the activity parameter, can be used to constrain the interior physics of neutron stars, and possibly also to determine the mass of glitching pulsars.
I have also discussed the vortex avalanche model for the trigger mechanism, which could explain the observed size and waiting time distributions in most pulsars, and finally discussed prescriptions to include sub-grid vortex motion in large scale hydrodynamical simulations.

In the near future, more detailed inputs from microphysics and constraints on the equation of state of dense matter (such as those obtained with the recent multi-messenger observations of a binary neutron star merger \citep{LIGO}) will increase the predictive power of our models, in order to make maximal use of new data and observations of glitches with instruments such as the Square Kilometre Array.

%\begin{thebibliography}{}  % PLEASE USE THE JOURNAL ABBREVIATIONS AS CAN BE FOUND IN THE LIST BELOW THE BIBLIOGRAPHY
\bibliographystyle{aasjournal}
\bibliography{glitches}

%\end{thebibliography}

\end{document}